\documentclass[conference]{llncs}
\usepackage{graphicx}
\usepackage{epstopdf}
\usepackage{url}
\usepackage{comment}
\usepackage{subfigure}

\newcommand\fixme[1]{}
\renewcommand{\qed}{\hfill \mbox{\raggedright \rule{0.1in}{0.1in}}}

\newbox\subfigbox 
\makeatletter
\newenvironment{subfloat}
{\def\caption##1{\gdef\subcapsave{\relax##1}}%
\let\subcapsave=\@empty 
\let\sf@oldlabel=\label
\def\label##1{\xdef\sublabsave{\noexpand\label{##1}}}%
\let\sublabsave\relax 
\setbox\subfigbox\hbox
\bgroup}
{\egroup 
\let\label=\sf@oldlabel
\subfigure[\subcapsave]{\box\subfigbox}}%
\makeatother

\begin{document}
\pagestyle{empty}
\title{Cloud and the City: Facilitating Flexible Access Control over Data Streams\thanks{This work has been supported by A*Star TSRP grant no. 1021580038 for `pCloud: Privacy in data value chains using peer-to-peer primitives' project. Contact author: Anwitaman Datta (anwitaman@ntu.edu.sg)}}
\author{Wen Qiang Wang\and Dinh Tien Tuan Anh \and Hock Beng Lim \and Anwitaman Datta}
\institute{Nanyang Technological University, Singapore}

\maketitle

\begin{abstract}
The proliferation of sensing devices create plethora of data-streams, which in turn can be harnessed to carry out sophisticated analytics to support various real-time applications and services as well as long-term planning, e.g., in the context of intelligent cities or smart homes to name a few prominent ones. A mature cloud infrastructure brings such a vision closer to reality than ever before. However, we believe that the ability for data-owners to flexibly and easily to control the granularity at which they share their data with other entities is very important - in making data owners feel comfortable to share to start with, and also to leverage on such fine-grained control to realize different business models or logics. In this paper, we explore some basic
operations to flexibly control the access on a data stream and
propose a framework \textit{eXACML+} that extends OASIS's XACML
model \cite{oasisxacml} to achieve the same. We develop a prototype
using the commercial StreamBase engine \cite{streambase} to demonstrate a seamless combination of stream data processing with (a small but important selected set of) fine-grained access control mechanisms, and study the framework's efficacy based on experiments in cloud like environments.
\end{abstract}

\section{Introduction}
Wide-scale deployments of sensors and smart mobile devices, as well as emerging technologies and trends such as Internet of Things (IoT) and participatory sensing - all envision developing interesting real-time data stream driven applications not only based on the isolated data-streams generated by individual sources, but also by mashing them up together to generate more sophisticated services. Such services vary in scale and scope - from smart homes to intelligent cities \footnote{\url{http://en.wikipedia.org/wiki/Intelligent_cities}}, which may fuse together data all owned by a single owner or from many parties. The pervasive cloud infrastructure is an important catalyst in enabling such visions, because (i) it is cost effective \cite{roxana2009} for individual data owners since it provides the necessary computational and storage resources elastically, (ii) it allows fast prototyping, deployment and testing of new applications and analytics at large-scale, and equally importantly, (iii) by naturally achieving collocation of data from different sources that may be needed to build such complex applications, it reduces the barriers of sharing and collaboration.

For example, a flu outbreak within one or several cities can be tracked, studied (and necessary intervention measures taken) using real-time data from say hospitals, transport departments, weather stations, as well as telecom companies. In the pre-cloud era, finding and sharing such data would encounter several layers of barriers.

However, in order to facilitate meaningful sharing of data from data-owners to various users, it is crucial that the data-owners have adequate controls on what they wish to share and with whom. Such controls are desirable for several reasons, including maintaining ownership of data, privacy of content selectively, as well as monetization of the data by differentiated pricing and exposing data in different details to different users. In absence of fine-grained control, or if the control comes only at prohibitive costs (either in negotiating, interpreting or implementing), then it is more than likely that the data-owners would just decide to not share. In our previous work \cite{anh12}, we have proposed some simple extensions to XACML \cite{oasisxacml} to represent and enforce fine-grained access controls (such as time-window based aggregation, trigger/threshold based access, etc.) on archived data, i.e., data stored in a relational database. In this paper, we look at the relatively harder problem on how to achieve similar fine-grained access control on data-streams.

Thus, while the sharing predicates we explore remain the same as in our previous work, the main challenges and contributions of the current work is precisely to deal with data-streams, which rely on fundamentally different set of technologies than data stored in a RDBMS. Developing access control model and mechanism on data streams is more challenging due to the characteristics of stream data and Data Stream Management System (DSMS).

- The data model used in DSMS are fundamentally different from those used in traditional RDBMS. DSMS deals with unbounded and fast changing time-series tuples in data streams. Access control enforcement, particularly when it is based on the content (such as, say a value based trigger or range predicate) is not a one-time operation, but a continuous procedure applied on the data streams. Whenever a new data tuple arrives, corresponding access control actions must be taken on it. Therefore, models and technologies developed for access control enforcement on RDBMS such as those in \cite{ferrari2000} cannot be readily adapted for DSMS.

- Temporal constraints, i.e., sliding windows, plays a crucial role in DSMS. In addition to normal constraints such as selection and projection, window-based aggregate operations into the access control model also needs to be considered.

In this work we propose the \textit{eXACML+} framework, by extending our previous work on fine-grained access control in RDBMS, namely eXACML \cite{anh12}, which in itself extends the popular XACML \cite{oasisxacml} standard. The \textit{eXACML+} framework adds fine-grained access control to a popular data stream management model, namely the Aurora model\cite{abadi03}. The main contributions of this paper are:


1. We extend XACML to enable fine-grained access control for \emph{continuous queries} \cite{babu2001}, i.e., standing queries which are continuously processing new incoming data streams. We express the fine-grain access control policies within \textit{obligations} blocks of XACML policies and make the Policy Enforcement Point (PEP) generate corresponding continuous queries from the obligations provides the Policy Decision Point (PDP) grant the access.\footnote{PEP and PDP are standard terms associated with XACML technology stack, and will be described in detail in next Section.} These continuous queries are expressed as query graphs and are sent to back-end DSMS for processing. We refer to this approach as \textit{XACML+}.

2. We integrate the various components in the \textit{eXACML+} framework. The framework consists of entities including data server, \textit{XACML+} instances, proxy server and client interface. Users send requests for data streams together with customized continuous queries and obtain stream handles, which point to the unique resource identifiers (URIs) of the processed data streams from the framework.

3. We show that customized queries issued by users, if not taken care of carefully, can give rise to information leak in case of sliding window policies (which our framework can detect and prevent). We also discuss possible improvements to the system efficiency by informing users of empty/partial results due to policy and query mismatches.

4. We instantiate the eXACML+ framework using Aurora's commercialized software StreamBase~\cite{streambase} as the back-end DSMS. We evaluate the performance of our prototype in a cloud-like environment. The results indicate that the framework incurs relatively constant overhead and is scalable.


The current work assumes a trusted cloud service provider, which itself has access to all the stored data, and furthermore honestly enforces the sharing constraints specified by the data-owners. Content confidentiality from the cloud service provider itself, and the service provider's accountability are also important and much researched topics, which are necessary extensions for the presented work but are out of the scope of this paper. 

The rest of the paper is organized as follows. Section \ref{sec:extend_xacml} describes details of XACML+,i.e.,
our extension to XACML to support stream data. Section \ref{sec:design}\ discusses the design of the \emph{eXACML+} framework. The prototype and evaluation of the framework are presented in Section \ref{sec:evaluation}. Section \ref{sec:relatedwork}\  discusses related work in the relevant research area and
finally we conclude our work and propose future work in Section \ref{sec:conclusion}.

\section{XACML policies for Stream Data}
\label{sec:extend_xacml}

\subsection{Overview of XACML and Aurora Model}
The \emph{eXtensible Access Control Markup Language} \cite{oasisxacml} is a
OASIS framework for specifying and enforcing access control. Policies are written in XML and
contain elements including \emph{subjects, resources ,
actions, obligations, etc.}. The framework consists two main components: a Policy
Decision Point (PDP) and a Policy Enforcement Point (PEP). The former
 manages policies and evaluate user requests against the stored policies, the result of
which are permit or deny decisions. PEP's main role is to marshall user requests and the
PDP results. In addition to permit/deny decision, the PDP also returns a set of
\emph{obligations} to the PEP. We extend this process for fine-grained access control by
embedding parts of the policies in obligations which are then processed by the PEP.

Aurora\cite{abadi03} is a popular model for stream data, which has matured
into a commercial product, i.e. the StreamBase engine\cite{streambase}. In this
model, a data stream consists of an append-only sequence of tuples with the same schema.
A query on a data stream is modelled as a directed acyclic graph,which we refer as \textit{query graph},  of operators (also called boxes). The query graph is applied to each tuple from the stream such each tuple in the output data stream satisfies all predicates in the query graph. StreamBase also comes with support for StreamSQL,which is SQL-like representation of query graphs.

The Aurora model supports a number of operators (or
boxes), but in this work we focus on three common ones: \emph{filter} (selection),
\emph{map} (projection) and \emph{window based aggregation} (aggregate functions applied
on sliding windows). A filter operator has a condition $C$ --- a boolean expression
composed of logic operators (\emph{AND}, \emph{OR}, \emph{NOT}), equality and inequality
operators ($<,>,\leq,\geq,=,\neq$). A map operator contains a set of projected attributes
$S$. A \textit{window-based aggregation} operator consists of the sliding window
(specified by the window type, size, advance step), the set of attributes and the
aggregate functions to be computed over each window.

\subsection{Fine-grained Access Control Policies}
\label{subsec:access_policies}
To better illustrate our access control model for stream data, we use the following
example throughout the paper.

\begin{example}
\label{exp:1}
The National Environmental Agency (NEA) wishes to provide real-time weather data service
through the Cloud platform. The data has the schema
(\emph{samplingtime, temperature, humidity, solar radiation, rain rate, wind speed, wind
direction, barometer}) and is generated every thirty seconds by a weather station. Instead
of creating one customised data stream for each individual customer, NEA decides to use
the cloud's access control mechanism. Benefits of this approach has been discussed in
\cite{anh12}. The Land Transport Authority (LTA) is developing an
automatic warning system that alerts drivers of possible traffic congestions due to heavy
rain. The warning system requires real-time weather data from NEA which specifies the
following policy: 1) only samplingtime, rain rate, and wind speed data are visible 2)
data should come in windows of size 5 and advance step of size 2, and the
functions applied on samplingtime, rain rate and wind
speed are \emph{lastValue}, \emph{average} and
\emph{maximum}, respectively 3) data is visible only when the rain rate is greater than
5mm/hour.  \end{example}

\begin{figure}
\centering
\includegraphics[scale=0.5]{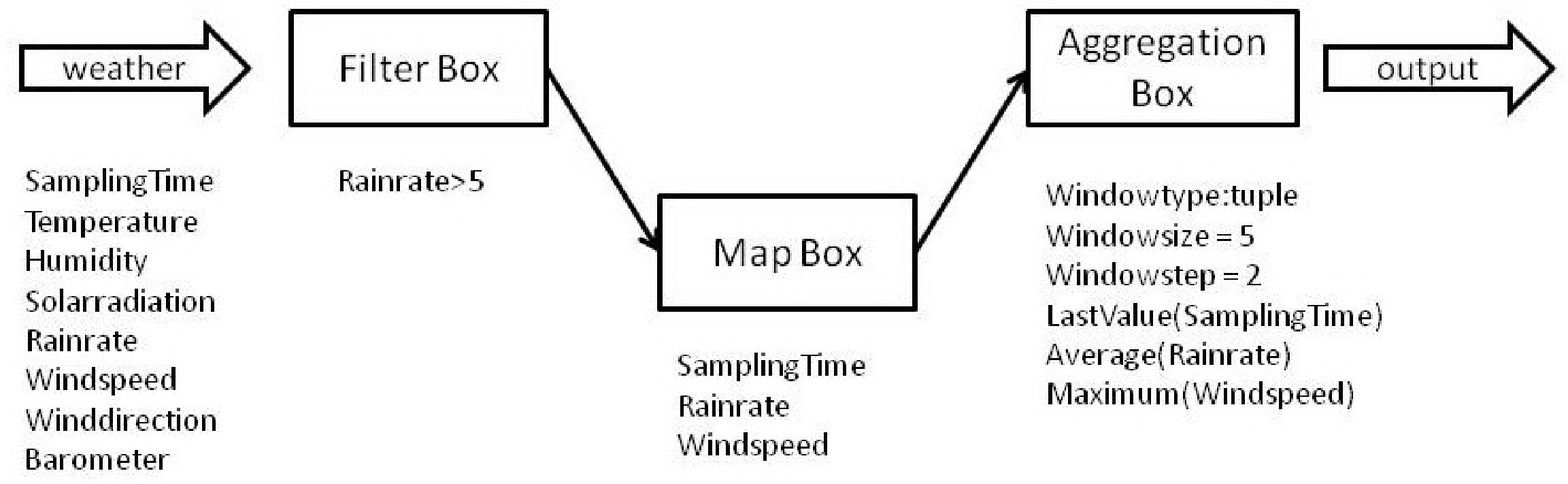}
\caption{Aurora query graph for the example in section \ref{subsec:access_policies}}
\label{fig:querynetwork}
\end{figure}

\begin{table}
\begin{tabular}{l|l}
\hline
\textbf{Description} & \textbf{Obligation Id} \\
\hline\hline
Filter & \textit{exacml:obligation:stream-filtering} \\
Map & \textit{exacml:obligation:stream-mapping} \\
Window-Based Aggregation & \textit{exacml:obligation:stream-window-aggregation} \\
\hline
\end{tabular}
\caption{Obligation types}
\label{tab:obligations}
\end{table}

\vspace{-11pt}

Figure~\ref{fig:querynetwork} shows the Aurora query graph that transforms the original weather data stream so that it satisfy the above access control scenario. Using the obligation-based approach
described in \cite{anh12}, we create new obligation elements, each for every operator (listed in
table~\ref{tab:obligations}). The detailed information of each obligation element is as follows:

1. Filter: consists of a string attribute with attribute ID:
\textit{exacml:obligation:stream-filter-condition-id}. The value is a string representing
a boolean expression used as the filter condition $C$.

2. Map: consists of a set of string attributes with ID:
\textit{exacml:obligation:stream-map-attribute-id}. The values are attribute names, used
used to restrict access to only authorized attributes, such as
rain rate and wind speed in the example.

3. Window-Based Aggregation: consists a number of attributes:

 - Window type: string attribute with attribute id:
\textit{exacml:obligation:stream-window-type-id}. It specifies if the window size is based
on number of tuples or number of time unit.

 - Window size: integer attribute with attribute id:
\textit{exacml:obligation:stream-window-size-id}. It specifies the size of the window in
the number of tuples or time units.

 - Window advance step: integer attribute with attribute id:
\textit{exacml:obligation:stream-window-step-id}. It specifies how fast the window
advances on the stream.

 - Aggregation attribute: string attribute with attribute id:
\textit{exacml:obligation:stream-window-attr-id}. It specifies the attribute in the data
stream schema that is to be aggregated, and what the aggregate function is. Its value is
 of the form \emph{ attribute-id:aggregate-function}, where\textit{
attribute-id} is the name of the attribute and \textit{aggregate-function} is an element from the set of aggregate functions \textit{\{Avg, Max, Min, Count, LastValue, FirstValue,...\}}.

Once we have defined the individual obligation elements, we can combine them to form the obligations block of the XACML policy for the access scenario. Figure~\ref{fig:xacmlPolicy} shows how these obligation elements are used in the policy.

\begin{figure}[h!]
\centering
\scriptsize
\begin{verbatim}
 <Obligations>
    <Obligation ObligationId="exacml:obligation:stream-filter" FulfillOn="Permit">
      <AttributeAssignment AttributeId="pCloud:obligation:stream-filter-condition-id"
      DataType="http://www.w3.org/2001/XMLSchema#string">rainrate > 5 </AttributeAssignment>
    </Obligation>
    <Obligation ObligationId="exacml:obligation:stream-map" FulfillOn="Permit">
          <AttributeAssignment AttributeId="pCloud:obligation:stream-map-attribute-id"
          DataType="http://www.w3.org/2001/XMLSchema#string">samplingtime</AttributeAssignment>
          <AttributeAssignment AttributeId="pCloud:obligation:stream-map-attribute-id"
          DataType="http://www.w3.org/2001/XMLSchema#string">rainrate</AttributeAssignment>
          <AttributeAssignment AttributeId="pCloud:obligation:stream-map-attribute-id"
           DataType="http://www.w3.org/2001/XMLSchema#string">windspeed</AttributeAssignment>
    </Obligation>
    <Obligation ObligationId="exacml:obligation:stream-window" FulfillOn="Permit">
      <AttributeAssignment AttributeId="pCloud:obligation:stream-window-step-id"
      DataType="http://www.w3.org/2001/XMLSchema#integer">2</AttributeAssignment>
      <AttributeAssignment AttributeId="pCloud:obligation:stream-window-size-id"
      DataType="http://www.w3.org/2001/XMLSchema#integer">5</AttributeAssignment>
      <AttributeAssignment AttributeId="pCloud:obligation:stream-window-type-id"
      DataType="http://www.w3.org/2001/XMLSchema#string">tuple</AttributeAssignment>
      <AttributeAssignment AttributeId="pCloud:obligation:stream-window-attr-id"
      DataType="http://www.w3.org/2001/XMLSchema#string">samplingtime:lastval</AttributeAssignment>
      <AttributeAssignment AttributeId="pCloud:obligation:stream-window-attr-id"
      DataType="http://www.w3.org/2001/XMLSchema#string">rainrate:avg</AttributeAssignment>
      <AttributeAssignment AttributeId="pCloud:obligation:stream-window-attr-id"
      DataType="http://www.w3.org/2001/XMLSchema#string">windspeed:max</AttributeAssignment>
    </Obligation>
 </Obligations>

\end{verbatim}
\caption{Obligation portion of the XACML policy for the example in section \ref{subsec:access_policies}}
\label{fig:xacmlPolicy}

\end{figure}
\vspace{-20pt}

\section{The eXACML+ Framework}
\label{sec:design}
\begin{figure}
\subfigure[eXACML+]{
\includegraphics[scale=0.3]{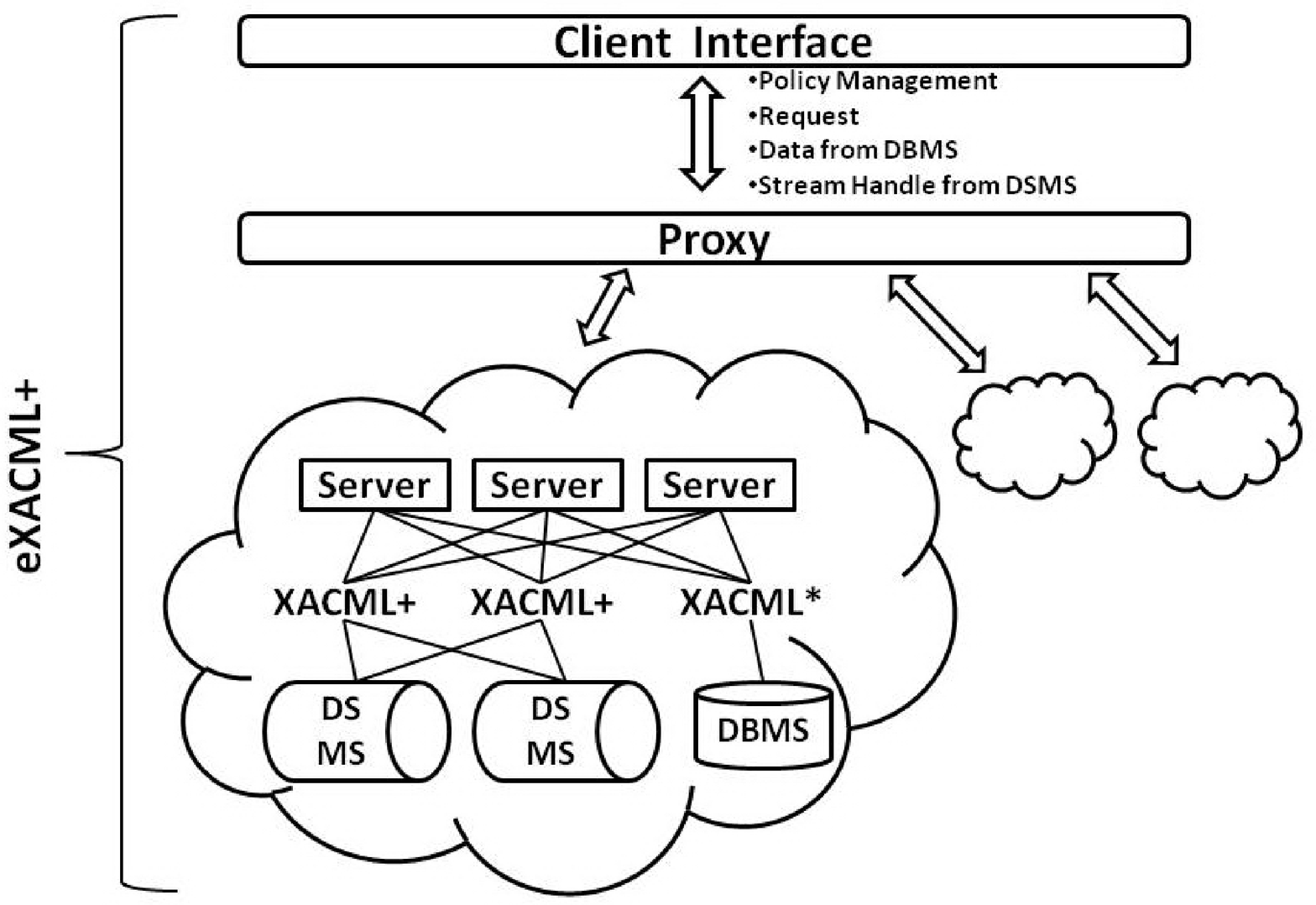}

\label{fig:exacml+}
}
\subfigure[XACML+]{
\includegraphics[scale=0.3]{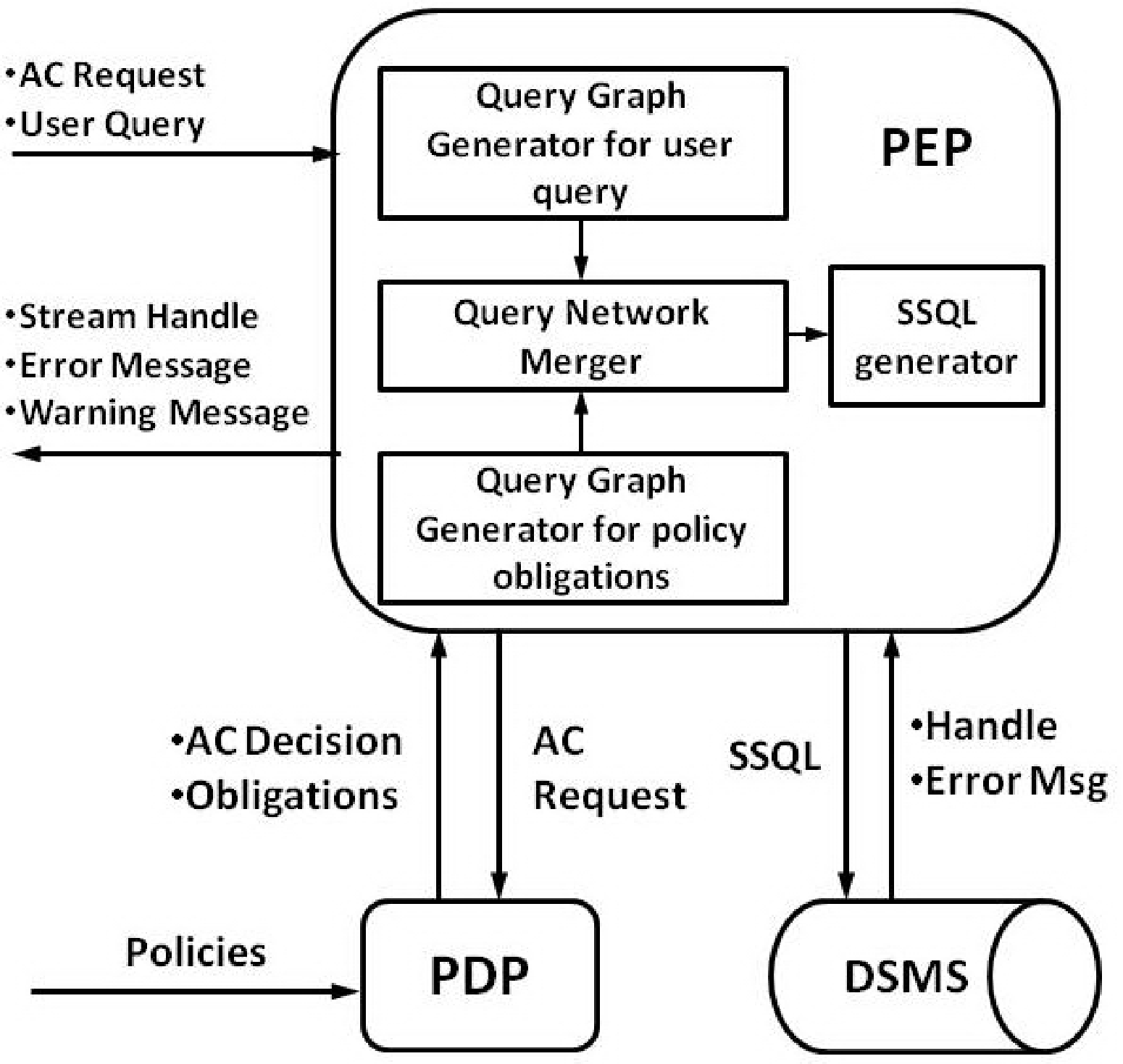}

\label{fig:extendedXACML}
}
\caption{Architecture of XACML+ and eXACML+ framework}
\end{figure}

We describe in this section the design of the \textit{eXACML+} framework which is a natural extension of the \textit{eXACML} with additional functionality to manage Aurora query graphs and customised queries issued by users. The architecture of \textit{eXACML+} resembles that of \textit{eXACML} except that: 1) New \textit{XACML+} instances are added into the framework to handle access control needs on data streams and 2)Update on the policy management module in response to the change of data model. Figure \ref{fig:exacml+} illustrates the architecture of eXACML+. It  includes entities such as \textit{cloud server},\textit{ XACML+} and\textit{ XACML*} instances, \textit{proxy} with cache feature and \textit{client} interface. We will also include discussion on two important issues which have not been discussed in the original eXACML framework,  which are 1) why allowing multiple windows on the same data stream could cause violation of data privacy and 2) how do we alert users if their queries contradict with the access control policies enforced on the data streams and could cause empty/partial result sets eventually.

\subsection{Handling User Query}
\label{subsec:userquery}
In many cases, the data stream accessible by the user may not directly fit the
actual requirement. In our example, suppose that the LTA later finds out that only rain
rate over 50mm/hr has influence on traffic condition, plus the warning system only need
the data from sliding windows of size 10 (instead of the original 5 tuples). The LTA could
process the incoming data steam locally. However, having such additional filtering done by
a server (the cloud) is more preferable. In our framework, the user sends a customised
query to the PEP. The query acts as a request to apply additional operation on the
authorized stream. We implement the query in XML form, as shown in Figure
\ref{fig:uq&ssql}(a). The PEP transforms this into an Aurora query graph
similar to that in Figure \ref{fig:querynetwork}, and then combines it with the query graph
derived from the policy obligations. One could simply concatenate the two graphs, but
properly merging them together gains advantages such as reducing the number of
operators in query graph and therefore improving efficiency. It also allows for detection
of empty/partial result (which we refer to as \textit{NR} and \textit{PR}).
\begin{figure}
\begin{subfloat}
\begin{minipage}{0.3\linewidth}
\scriptsize
\begin{verbatim}

<UserQuery>
   <Stream name="weather" />
   <Filter>
      <FilterCondition>
      RainRate > 50
      </FilterCondition>
   </Filter>
   <Map>
      <Attribute>RainRate</Attribute>
   </Map>
   <Aggregation>
      <WindowType>tuple</WindowType>
      <WindowSize>10<WindowSize>
      <WindowStep>2<WindowStep>
      <Attribute>avg(RainRate)</Attribute>
   </Aggregation>
</UserQuery>


\end{verbatim}

\end{minipage}
\caption{User Query in XML}
\label{fig:uq}
\end{subfloat}
\qquad
\qquad
\qquad
\begin{subfloat}
\begin{minipage}{0.3\linewidth}
\scriptsize
\begin{verbatim}
CREATE INPUT STREAM weather (
samplingtime timestamp , temperature double ,
 humidity double , rainrate double ,
 windspeed double,winddirection int,
 barometer double );

CREATE STREAM internal_0;
SELECT * FROM weather WHERE rainrate > 50 INTO internal_0;

CREATE OUTPUT STREAM internal_1;
SELECT internal_0.samplingtime,internal_0.rainrate,
FROM internal_0 INTO internal_1;

CREATE OUTPUT STREAM output;
CREATE WINDOW _10tuple( SIZE 10 ADVANCE 2 TUPLES);
SELECT lastval(samplingtime) AS lastvalsamplingtime,
avg(rainrate) AS avgrainrate
FROM internal_1[_10tuple] INTO output;

\end{verbatim}
\caption{StreamSQL statements}
\label{fig:ssql}
\end{minipage}

\end{subfloat}
\caption{User Query and StreamSQL}
\label{fig:uq&ssql}
\end{figure}
Merging two query graphs is equivalent to merging each type of operators in
the graphs. We explain how \textit{NR} and \textit{PR} can be detected during the merging process later in Section
\ref{subsec:nr/pr_alert}.  The rules for merging individual types of operators are:

 - Two filter operators $F_{1}$ and $F_{2}$ with condition $C_{1}$ and
$C_{2}$ are merged into a filter $F_3$ with the  condition $C_{3}$ = ($C_{1}$) AND
($C_{2}$). There are
cases that $C_{3}$ can be further simplified. For example, if $C_{1}$ = x $>$ $v_{1}$ and
C2 = x $>$ $v_{2}$, C3 can be written as $x > v_{2}$ iff $v_{2}$ $\geq$ $v_{1}$.

 - Two map operator $M_{1}$ and $M_{2}$ with attribute sets $S_{1}$ and $S_{2}$ are
merged into new operator $M_3$ with the attribute set $S_{3} = S_1 \cup S_2$.

.- Two window-based aggregation operator $A_{1}$ and $A_{2}$ are merged only if the
following conditions are met: 1) window types are the same 2)
suppose $A_{1}$ is derived the policy obligations and $A_{2}$ from user query,  $A_{1}$'s
window size and advance step are must be less than or equal to those of $A_{2}$. The
second condition is to ensure that user are not given more fine grained data than
permitted by the policy. The new operator $A_{3}$ will have the same window type as
$A_{1}$ and $A_{2}$, and the window size and advance step are the same as those of
$A_{2}$. The aggregation function and attribute sets are the intersection of those from
$A_{1}$ and $A_{2}$.

Figure \ref{fig:uq&ssql}(b) shows the StreamSQL statements after merging the query graph in
Figure \ref{fig:querynetwork} with user query in Figure \ref{fig:uq&ssql}(a).

\subsection{Design of XACML+}
\label{subsec:framework}

Figure \ref{fig:extendedXACML} shows the design of XACML+, which is an extension of the original oasis XACML model \cite{oasisxacml}. The work-flow is as follows:

1. PEP receives a user's request for accessing a stream, together with a customized
query, which are then forwarded to the PDP. The customised query is also converted into a
Aurora query graph.

2. PDP evaluates the request against the stream's policies and returns the decision and
obligations (if any) to the PEP. If the decision is Permit, PEP will generate a query
graph from the obligations.

3. PEP checks that for the credentials included the request, no query is
currently being applied to the same data stream. The reason for this is
given in Section \ref{subsec:multiple_windows}.

4. PEP merges the two query graphs derived from obligations and user query, during
which PR or NR are checked.

5. If there is no PR or NR warning detected, the merged query graph is converted into a StreamSQL
script and sent to the  data stream engine. A handle, in the URI form, is returned to
the user.

\subsection{Query Graph Management}
\label{subsec:policymanagement}

In the original eXACML framework that handles only bounded data, PDP is always called whenever a data request is received and only when the decision is 'Permit', SQL queries are generated from obligations and sent to database to retrieve data. This work flow guarantees that removing or updating a policy does not affect the privacy of the data owner. However, this is not the case when dealing with unbounded stream data. Instead of actual data, only a handle used to retrieve the actual data stream from data stream engine is returned as the response to user's request. The user then use this handle to connect to the back-end data stream engine for data streams. If the data stream owner for some reason has removed or modified the policy that grants the user for a particular data stream, the user may still connected to the data stream though he is not supposed to be able to access the data stream any longer.

To solve this issue, we need to in-cooperate query graph management into the framework. In additional to keeping track of policies loaded, data server also keeps track of query graphs that are generated by PEP and have already been sent to back-end data stream engines. In current eXACML+ design, whenever a policy has been removed or modified by user, all query graphs that are spawned by the policy are immediately withdrawn from back-end data stream engines. This may not be a flexible solution, but it ensures data privacy and is not easy to be compromised.

\subsection{Multiple Aggregation Windows}
\label{subsec:multiple_windows}

As described in Section \ref{subsec:framework}, only a single access is permitted on a
particular data stream for one user at any time. We justify this constraint by showing a
example in which one can reconstruct the raw data steam by combining outputs from multiple
aggregation windows of different window sizes or advance steps.

\begin{example}
\label{exp:2}

Suppose we have a single-attribute stream  \textit{S} =
$a_{0}$,$a_{1}$,$a_{2}$,$a_{3}$ ,$a_{4}$,$a_{5}$..., $a_{n}$. The access control policy
for $S$ allows for aggregation window $w$, where \textit{w.size} = 3,
\textit{w.advancestep} = 2, \textit{w.type} =\textit{ tuple},
\textit{w.attribute}=\textit{a} and\textit{ w.function}
=\textit{sum}. An user can request an aggregation window $v$ using a customised query,
provided that \textit{v.size}$>=$\textit{w.size},\textit{v.advancestep} $>=$
\textit{w.advancestep},
\textit{v.type} = \textit{w.type}, \textit{v.attribute} = \textit{w.attribute} and
\textit{v.function} = \textit{w.function}. If multiple accesses are allowed, the
user can obtain multiple result stream using different $v$ simultaneously. Let
$v_{1}$.\textit{size} = 3, $v_{2}$.\textit{size} = 4, $v_{3}$.\textit{size} = 5 and all
other window specifications be identical to \textit{w},
the framework will return the user three aggregated data streams $S_{1}$, $S_{2}$ and
$S_{3}$, such that:

\noindent$S_{1}$ = ($a_{0}$+$a_{1}$+$a_{2}$), ($a_{2}$+$a_{3}$+$a_{4}$),($a_{4}$+$a_{5}$+$a_{6}$), ... \newline
$S_{2}$ = ($a_{0}$+$a_{1}$+$a_{2}$+$a_{3}$),($a_{2}$+$a_{3}$+$a_{4}$+$a_{5}$),($a_{4}$+$a_{5}$+$a_{6}$+$a_{7}$),... \newline
$S_{3}$ =  ($a_{0}$+$a_{1}$+$a_{2}$+$a_{3}$+$a_{4}$),($a_{2}$+$a_{3}$+$a_{4}$+$a_{5}$+$a_{6}$), ... \newline

By computing $S_{2}$ - $S_{1}$ and  $S_{3}$ - $S_{2}$ , we obtain two new streams
$S^{'}$= $a_{3}$,$a_{5}$,$a_{7}$,... and $S^{''}$ = $a_{4}$,$a_{6}$,$a_{8}$,...
Merging $S^{'}$ and $S^{''}$, we can reconstruct all of the original stream except for
first three tuples.

\end{example}

In general, given a set of result streams derived from different aggregation windows
 with fixed advance step \textit{M} and different window size \textit{N, N+1, N+2,..,
N+M}, we can reconstruct the original stream from the \textit{Nth} tuple. The inductive proof is as following:

Suppose we have three $sum$ aggregation windows with sizes $N$, $N+Q_1$, $N+Q_2$, where $Q_1 < Q_2$, and a fix step size $M$, the first $k$ tuples of the three streams are: $S_0$ =($a_0$ + ... + $a_{N-1}$),($a_M$ + ... +$a_{N+M-1}$),...,($a_{kM}$ + ... +$a_{N+kM-1}$); $S_1$ =($a_0$ + ... + $a_{N+Q_1-1}$),($a_M$ + ... +$a_{N+M+Q_1-1}$),...,($a_{kM}$ + ... +$a_{N+kM+Q_1-1}$);$S_2$ =($a_0$ + ... + $a_{N+Q_2-1}$),($a_M$ + ... +$a_{N+M+Q_2-1}$),...,($a_{kM}$ + ... +$a_{N+kM+Q_2-1}$). Let $T_1$ = $S_1$ - $S_0$ and $T_2$ = $S_2$ - $S_1$, we can have $T_1$=($a_N$ +...+$a_{N+Q_1-1}$),..., ($a_{N+kM}$+...+$a_{N+kM+Q_1-1}$) and $T_2$=($a_{N+Q_1}$ +...+$a_{N+Q_2-1}$),..., ($a_{N+kM+Q_1}$+...+$a_{N+kM+Q_2-1}$).

In a similar fashion, we can construct subsequence streams until $T_{i}$ such that $T_{i}$=($a_{N+Q_{i-1}}$+...+$a_{N+Q_{j-1}}$), ...,($a_{N+kM+Q_{i-1}}$ + ... + $a_{N+kM+Q_{j-1}}$). Let $Q_1$=1 ,$Q_j$=$Q_i$+1, i.e., each window contains one more element than previous windows, and $Q_j<M$, we can simplify $T_1$ to $T_i$ as: $T_1$=$a_{N}$,$a_{N+M}$,...,$a_{N+kM}$, $T_2$=$a_{N+1}$,$a_{N+M+1}$,...,$a_{N+kM+1}$,...,$T_M$=$a_{N+M-1}$,$a_{N+2M-1}$,...,$a_{N+(k+1)M-1}$. Combine $T_1$ to $T_M$ in an interleave manner, we can obtain stream $a_N$,$a_{N+1}$,...,$a_{N+(k+1)M-1}$, which is the original stream except for first (N-1) tuples. \qed

\subsection{Checking for Empty or Partial Result Set (NR/PR)}
\label{subsec:nr/pr_alert}

As mentioned earlier, \textit{NR} and \textit{PR} warnings occur while the framework merge two
query graphs. Since all query graphs in our context are built with filter, map and window
based aggregation operators, we simplify this problem by checking how merging individual
operators causes \textit{NR/PR} cases. Let us first present the detailed definitions of \textit{NR} and \textit{PR} and an example of \textit{NR/PR} cases.

\textit{Partial Result Warning(PR)}: A system issued warning, stating that some tuples in the requested stream may not be returned to the user due to conflict between the user query and some policies enforced on the streams.  

\textit{Empty Result Warning(NR)}: A system issued warning, stating that none of the tuples in the request stream will be returned to the user due to conflict between the user query and some policies enforced on the streams. This must be differed from the case where the user does not have access to the stream. 

\begin{example}

Suppose we have a stream $S$ with single attribute {a} and filter condition $F1$: $a > 8$ from policy obligation, filter condition $F2$: $a > 5$ is from user query. Let a part of $S$ be (..., 9,10,11,3,2,6,9,8,7,2,13,...), the user query expects output to be like, (...,9,10,11,6,9,8,7,13,...). However, due to $F1$, tuples like {6},{8},{7} are filtered out and the actual stream the user will get is  (... 9,10,11,9,13,...). In this case, a $PR$ warning will be issued to the user stating that there are possibilities that some tuples that fit his requirement are not returned to him due to certain access control policies enforced on the data stream. If we change $F1$ to be $a < 4$, only {3,2,2} will be retained after applying $F1$ on $S$. Clearly, any real number less than 4 is never greater than 5, so none of the tuples will make $a > 5$ true. In another word, predicate ``$a<4$ AND $a>5$'' will always be false no matter what value $a$ may take. Therefore, none of the tuples will be returned to the user and a $NR$ warning will be issued.

\end{example}

Now let us look at how to generate \textit{NR/PR} warning for each operator:

\textit{Map Operator}: Suppose we have two map operators: $M_{1}$ from policy and $M_{2}$ from user query with attribute sets $S_1$ and $S_2$. If $S_1 \cap S_2 = \emptyset$, alert \emph{NR}. Otherwise, alert \emph{PR} if $S_1 \ne S_2$.

\textit{Aggregate Operator}: Suppose we have two aggregation operators $A_{1} $ and $A_{2}$, where $A_{1}$ comes
from the policy and $A_{2}$ comes from user query.We apply following rules: (1)If  $A_{1}.size > A_2.size$, alert \textit{NR};(2)If  $A_1.advancestep > A_2.advancestep$, alert \textit{NR};(3)If  $A_1.type \neq A_2.type$, alert \textit{NR};(4)If different aggregation functions are applied for the same attribute in $A_1$ and
$A_2$, alert \textit{NR};(5)For every attribute \textit{a} in $A_{2}$, if \textit{a} is an attribute in $A_{1}$
and the aggregation functions applied to \textit{a}  in both $A_{1}$ and $A_{2}$ are the
same, do not alert;(6)In all other cases, alert \textit{PR}.

\textit{Filter Operator}:
Checking if merging two filter operators gives rise to \textit{NR} or \textit{PR} warnings is more
complicated. We first define two terms: (1)a \textit{simple expression} $S$ is an expression of the form ``$x$ $op$ $v$'', where x is a variable(in our case, an attribute name of a stream schema), $op$ $\in$ \{$<$,$>$,$\geq$,$\leq$,$=$,$\neq$\},and $v$ is a number, or a string (only when $op$ is = or $\neq$). (2)A \textit{complex expression} $C$ is an logical predicate that is formed by connecting \textit{simple expressions} with \textit{NOT}, \textit{OR} or \textit{AND}. In our case, $C$ is the filter condition of a filter operator that belongs to a policy or a user query. 

Suppose we have two filter operators $F_{1}$ and $F_{2}$ with condition $C_1$
and $C_2$ respectively. Following procedure is used to do the check:

\begin{table}
\centering
\begin{tabular}{l|l}
\hline
\textbf{\textit{x op v}} & \textbf{\textit{x op' v}} \\
\hline\hline
x $>$ v& x $\leq$ v \\
x $<$ v& x $\geq$ v \\
x $\geq$ v& x $<$ v \\
x $\leq$ v& x $>$ v \\
x $=$ v& x $\neq$ v \\
x $\neq$ v& x $=$ v \\

\hline
\end{tabular}
\caption{Rules to Convert \textit{NOT (x op v ) }to \textit{x $op^{'}$ v}}
\label{tab:not_convertion}
\vspace{-20pt}
\end{table}

\textit{Step 1:} Let $P$ = $C_1$ AND $C_2$. Eliminate \textit{NOT} operator in $P$ using De Morgan's laws and rules listed in table \ref{tab:not_convertion}. Let the result expression be $P_{1}$.

\textit{Step 2:} Convert $P_{1}$ into its disjunctive normal form (DNF) $P_{2}$. Note each variable $S$ in $P_{2}$ is a \textit{simple expression}. The conversion is done by first change $P_{1}$ to its postfix form and then evaluate the postfix expression, the algorithms of which are standard stack based algorithms and can be found easily in places such as in \cite{postfix}. While doing the postfix evaluation, if the operator is $AND$, apply distribution law on the two operands. If the operator is $OR$ , concatenate two operands with OR.

\textit{Step 3:} Check for \textit{NR} and \textit{PR} in $P_{2}$, by pair-wisely
calling function \textit{checkTwoSimpleExpression} on each two $S$ within the same conjunctive expression. If any function call returns with NR or PR, mark the whole conjunctive expression with NR or PR. If all conjunctive expressions  are marked with PR or NR, alert
PR or NR, respectively. The cost of the whole procedure is bound by O($kn^{2}$),where $k$ is the number of conjunctive expressions in $P_{2}$ and $n$ is the maximum number of $S$ in a single conjunctive expression. 

\begin{figure}
\includegraphics[scale=0.5]{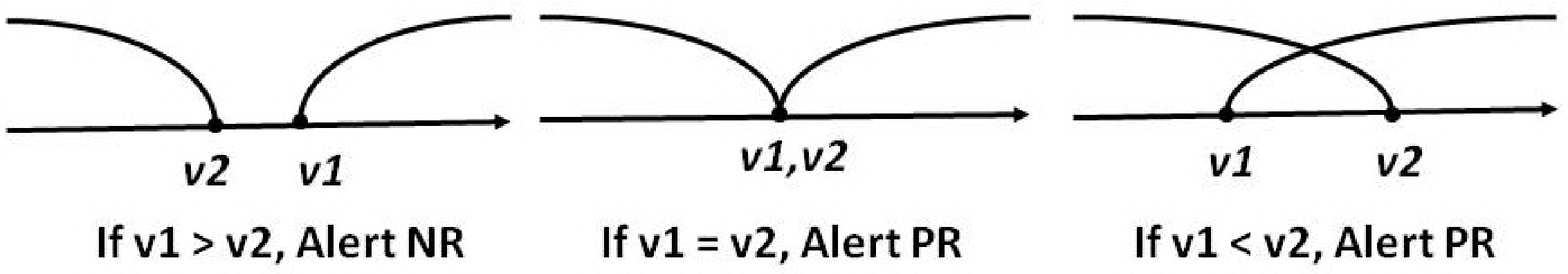}
\caption{Checking PR/NR for $S_{1}$ = x $\geq$ $v_{1}$ and $S_{1}$ = x $\leq$ $v_{2}$}
\label{fig:mergeexamle}
\end{figure}

The function \textit{checkTwoSimpleExpression} takes as inputs two simple expressions
$S_{1}$ and $S_{2}$. Apparently, checking is only necessary when $S_{1}$.x = $S_2$.x. Since there are six possible values of $op$ $\in$ \{$<, >, \leq, \geq, =, \neq$\}, we need
to do $6^{2}$ comparisons to include all cases that $op$ may take in $S_1$ and $S_2$. For each comparison, there are three cases, i.e., $S_{1}$.v $>$ $S_{2}$.v,  $S_{1}$.v $<$ $S_{2}$.v and $S_{1}$.v $=$
$S_{2}$.v. We show here an
example of how to generate $NR$ and $PR$ alerts for one comparison. Let $S_{1}$ = x $\geq$
$v_{1}$ and $S_{2}$ = x $\leq$ $v_{2}$, Figure
\ref{fig:mergeexamle} shows how the warnings are produced given different values of
$v_{1}$ and  $v_{2}$ and Example \ref{exp:nr/pr} better illustrates how the whole procedure works.

\begin{example}

Suppose we have $C_1$ =(a$>$20 AND a$<$30) OR NOT(a$\neq$40), $C_2$ = NOT(a$\geq$10) AND b=20. 

First, we eliminate $NOT$ in $P$ = $C_1$ AND $C_2$ as described in Step 1. After elimination, we have $P_1$ = (a$>$20 AND a$<$30) OR a$=$40 AND a$<$10 AND b=20. To make it easier to read, we use following substitutions: Let $A$ be $a>20$, $B$ be $a<30$, $C$ be $a=40$, $D$ be $a<10$, $E$ be $b=20$, $\&$ be AND, $\|$ be OR. $P_1$ is then changed to ((A$\&$B)$\|$C)$\&$(D$\&$E) 

Then we convert $P_1$ to its postfix form, which is A B $\&$ C $\|$ D E $\&$ $\&$, and evaluate it as described in Step 2. After evaluation, we will have $P_2$= E $\&$ D $\&$ C $\|$ E $\&$ D $\&$ B $\&$ A, which is in CNF and has identical truth table as $P_1$.

In Step 3, we pair-wisely apply function  \textit{checkTwoSimpleExpression} on simple expressions of the two conjunctive expressions in $P_2$, which are $e_1$ = E $\&$ D $\&$ C and $e_2$ = E $\&$ D $\&$ B $\&$ A. For $e_1$, $C_3^2$ = 3 calls are made on simple expression pairs (E,D), (E,C) and (D,C). For $e_2$, $C_4^2$ = 6 calls are made on (E,D),(E,B),(E,A),(D,B),(D,A) and (B,A). Among all these function calls, (D,C) returns $NR$ as a<10 and a=40 cannot be true for any given $a$ value. Similarly, function calls on (D,A) also returns $NR$ as a$<$10 and a$>$20 contradicts. Both $e_1$ and $e_2$ cannot be true for any $a$ value, means that $e_1\|e_2$, whose truth table is identical to $P_1$ and $P_2$, cannot be true for any $a$ value. In this case, a $NR$ warning will be returned to the user, stating that some predicates in his query contradicts with some predicates in the police and no result will be produced even if he has been granted access to the target stream.

\label{exp:nr/pr}
\end{example}

\section{Prototype and Evaluation}
\label{sec:evaluation}
\subsection{Prototype Implementation}
\label{subsec:implementation}
We have implemented a prototype of \textit{eXACML+} in Java. We use APIs provided by StreamBase to manage and query data streams. Communications between clients, proxies and servers are socket-based. For XACML+, we extended sun's XACML implementation \cite{sunxacml}
, which is an open source Java project for the XACML standard. The prototype supports access control over data streams by dynamic managing query graphs generated from user query and policies.

\subsection{System Evaluation}
\label{subsec:evaluation}
The performance of the system is measured by the time taken to fulfil user's requests on data streams.We compare the results with that of a system that query directly to StreamBase DSMS, which is refer to as \textit{direct-query} system.

\textbf{Hardware Setup:} The prototype system is deployed in a cloud-like environment. We make use of four machines in the experiments, which run data server, StreamBase, proxy and client interface accordingly. Machines running server and StreamBase are both IBM x3650 servers located in a server room and each has two Quad-Core Intel Xeon E5450 processors and 32GB memory. The machine running proxy is a mini work station and the client machine is one author's macbook. All machines are connected via University's 100Mbps Intranet. The StreamBase DSMS maintains a few real-time data streams from various projects, such as weather data feeds from a number of mini weather stations producing weather records at one minute interval.There are also GPS track information from personal mobile devices.

\textbf{Workloads:} Workloads are formed by sequences of continuous queries. Each continuous query corresponds to three files in the experiment: (1)a StreamSQL script as the input to direct-query system;(2)a XACML policy file whose obligations forms the query graph exactly as that in the above StreamSQL script. This file will be loaded into eXACML+ to provide access control policies to PDP;(3)a XACML request file for requesting data streams from eXACML+, which may also have user query embedded inside. The request file contains credentials, resources and actions as specified in the corresponding policy so that PDP will always permit the request so that PEP can generate query graphs from obligations and user queries. The actual specifications of each query graph are generated randomly, but we make sure that parameter names are consistent with those in stream schemas of so that every query graph generated from PEP is valid. In this experiment, we form query graphs with a pre-defined set of combinations for Filter(\textit{FB}), Map(\textit{MB}) and Aggregation (\textit{AB}) operators. The sequence of continuous queries follow two set up: (1) Each continuous query and corresponding request appears only once (i.e. is unique) in the sequence;(2)The sequence follows Zipf distribution, which models the scenario where a small number of popular streams are requested frequently. Such request pattern is popular in P2P file-sharings and web caching \cite{adamic02,klemm04} and we use it to verify the performance improvement brought by cache mechanism on query graphs in proxy. The parameters used to generate the workloads are illustrated in Table \ref{tab:variables} .

\begin{table*}
\centering
\scriptsize
\begin{tabular}{|l|l|p{6cm}|}
  \hline
  Variable & Value & Description\\ \hline

  $nDirectQueries$ & 1500 & number of direct queries \\ \hline

  $directQueryDist$ & 160:170:130:124:254:290:372 & query graph composition
   (Single FB : Single MB : Single AB : FB+MB : FB+AB : MB+AB : FB+MB+AB) \\ \hline

  $nPolicies$ & 1000 & number of unique policies \\ \hline

  $nRequests$ & 1500 & number of matching requests \\ \hline


  $\alpha$ & 0.223 & skew parameter for Zipf distribution\\ \hline

  $maxRank$ & 300 & maximum rank of unique requests from which Zipf
  distribution is generated \\ \hline
\end{tabular}
\caption{Summary of parameters used experiments}
\label{tab:variables}
\vspace{-20pt}
\end{table*}

\begin{figure}

\subfigure[Unique query \& request sequence ]{
\hspace{-4.5em}
\includegraphics[scale=0.55]{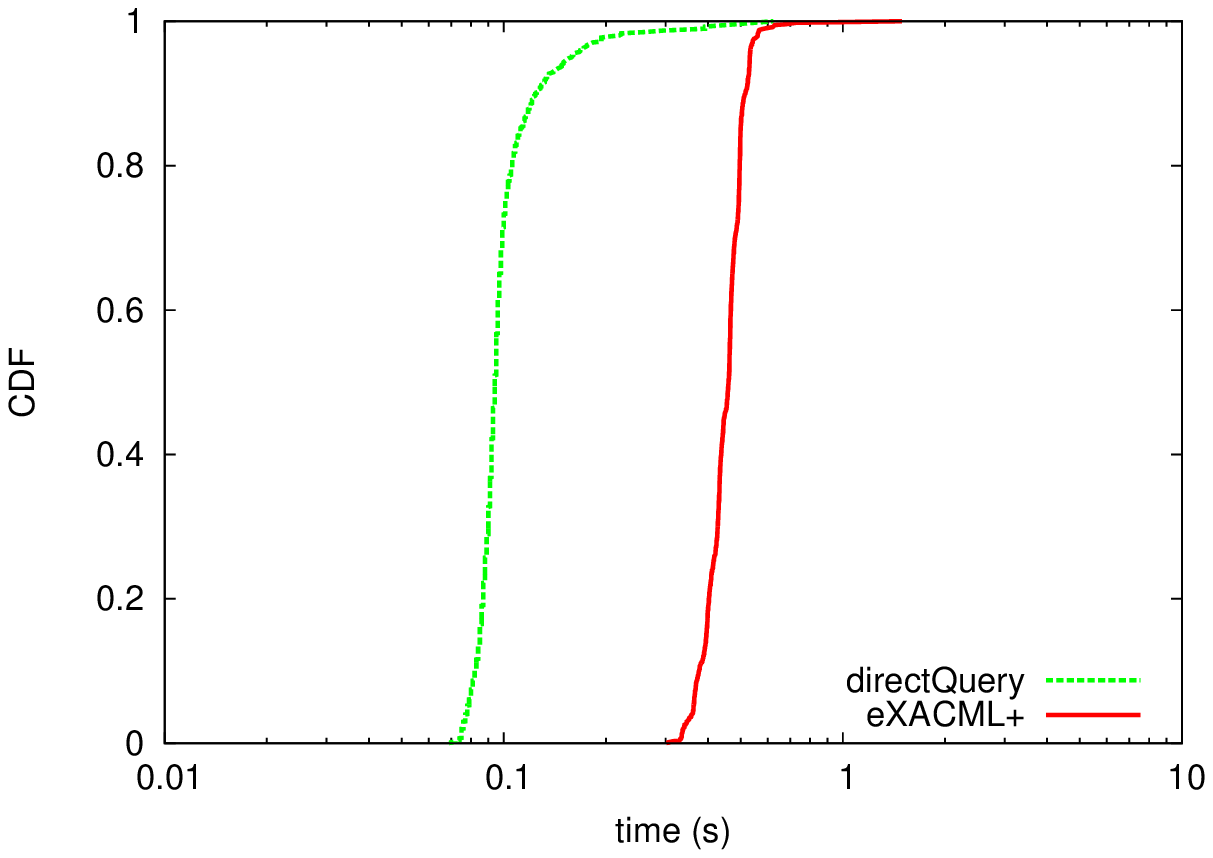}
\label{fig:unique_dis}
}
\subfigure[Zipf distributed sequence]{
\includegraphics[scale=0.55]{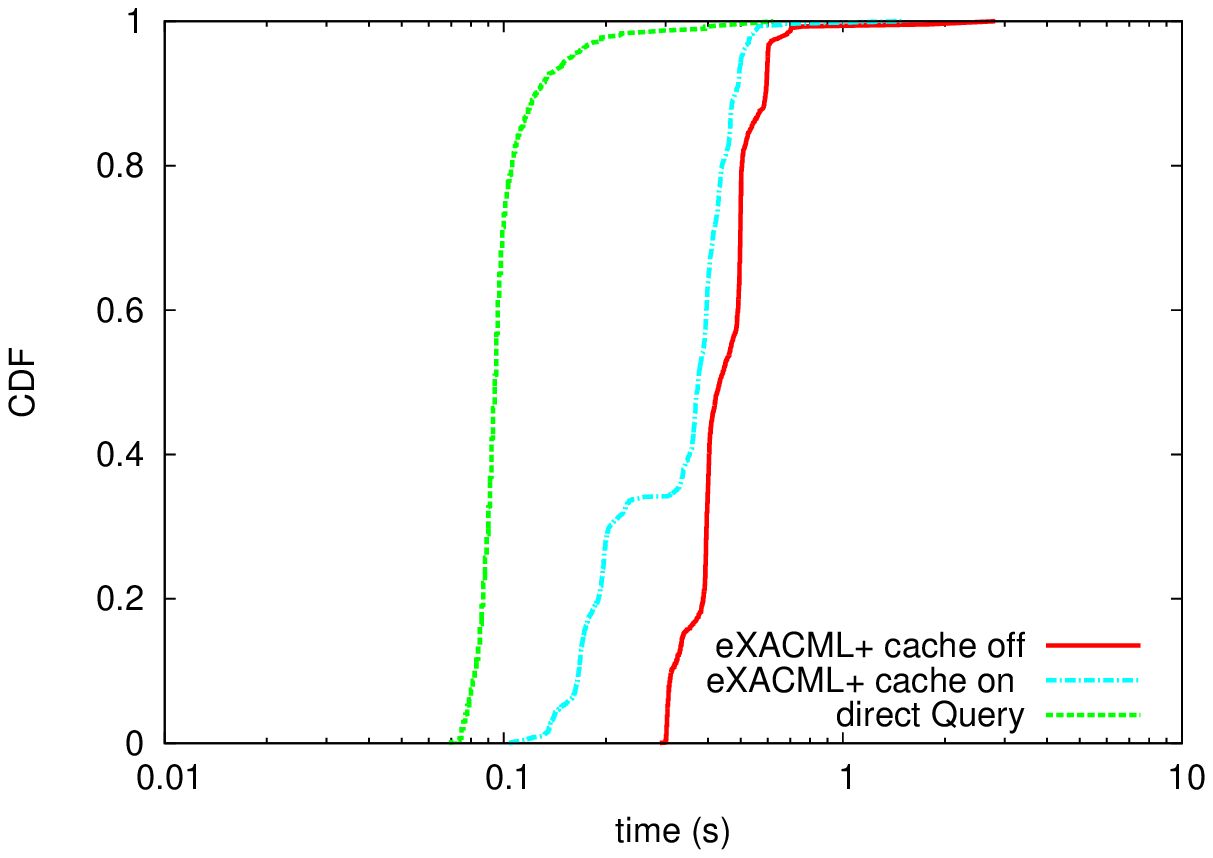}
\label{fig:zipf_dis}
}
\label{fig:experiment_1}
\caption{Overall Performance}

\end{figure}

\textbf{Experiments}: We measure system performance in terms of time taken to fulfil authorized data stream requests. Apparently, access control frameworks incur additional overhead when handling request from users.  However, as we will show in the results, the overhead of the framework is consistent most of the time and cache mechanism implemented in proxy does help improve the overall performance. Before any user request is made, we need to load policies onto the data servers so that PDP can make decisions based on them. Loading a policy onto server takes a small amount of time without respect to the number of policies already loaded. The average loading time is 0.25 second with standard deviation of 0.06 second.

\begin{figure}

\subfigure[100 requests]{
\hspace{-4.5em}
\includegraphics[scale=0.55]{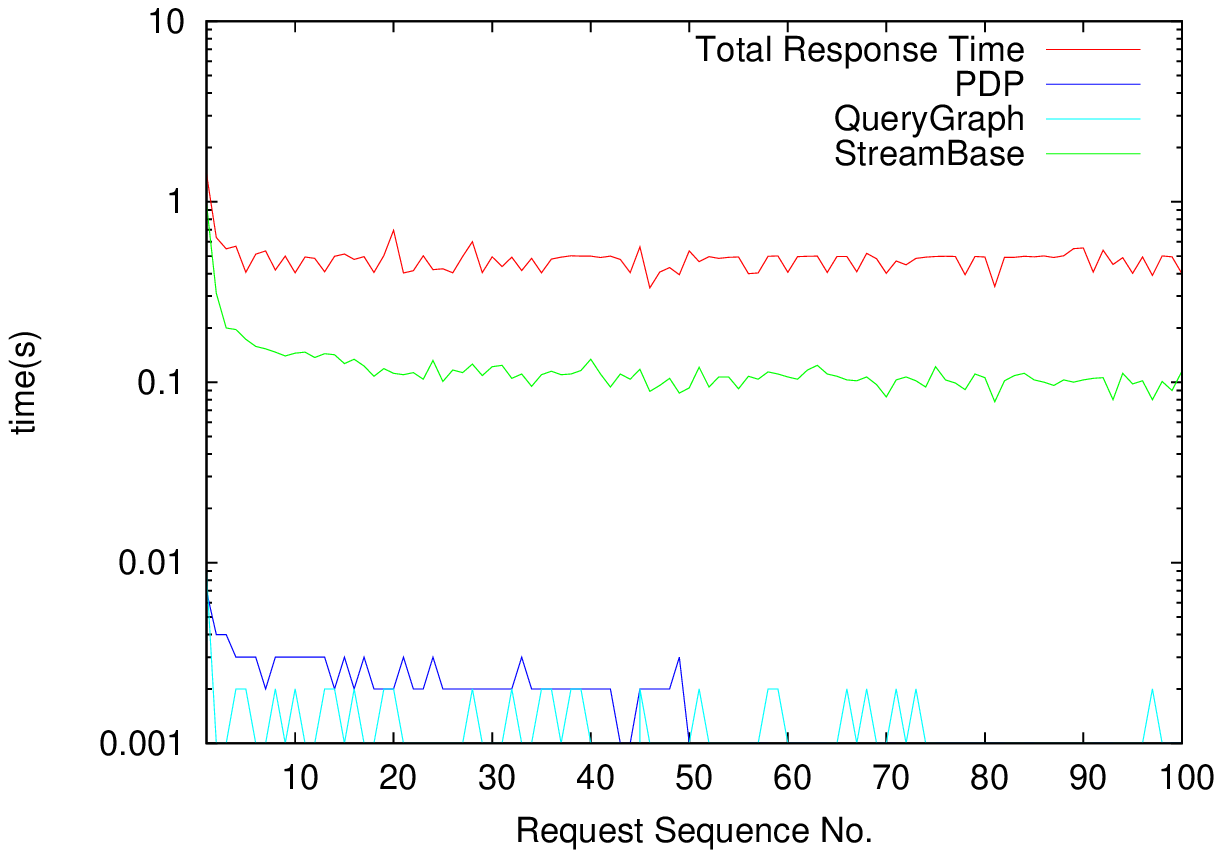}
\label{fig:breakdown_100}
}
\subfigure[1500 requests]{
\includegraphics[scale=0.55]{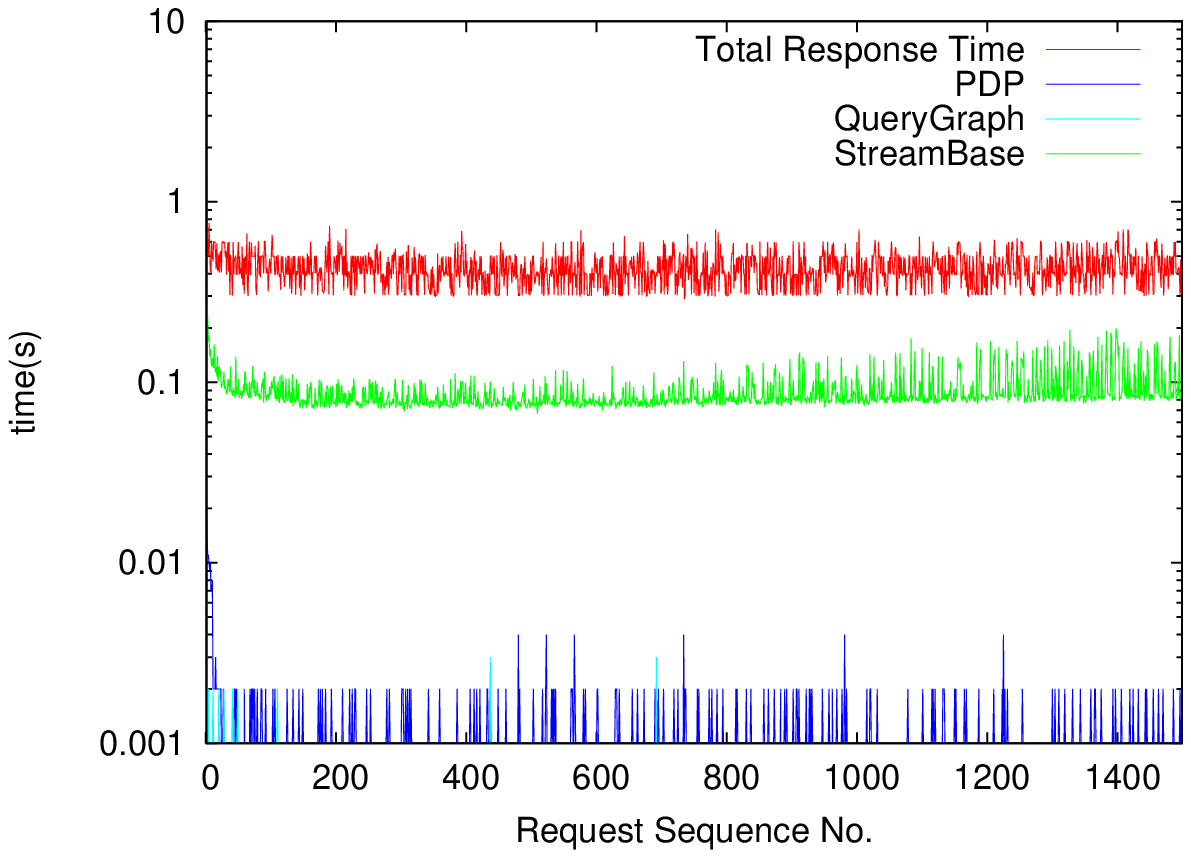}
\label{fig:breakdown_1500}
}
\label{fig:experiment_2}
\caption{Detailed Processing Time of AC Requests}

\end{figure}

Figure \ref{fig:unique_dis} presents the result obtained by running the unique query/request sequence. Different from \textit{eXACML}, there are no actual data transferred in the system and most direct queries and most requests to \textit{eXACML+} can get response very quickly,i.e.,less than one second. The response time for direct query is consistent most of the time, which is reasonable as only the query graphs,in forms of StreamSQL scripts, and data stream handles, in the form of URIs, are exchanged between the direct query system and DSMS. Given a constant environment in terms of network condition, computational power, etc., the response time should not vary much. \textit{eXACML+} incurs overheads and the curve is less smooth compared to that of direct query system. We believe it is mainly because of additional network traffic among clients, proxies and servers, which occupies about two thirds of the total response time. The cost for communication between multiple entities is also less predictive and subject to charge with large variance. Figure \ref{fig:breakdown_100} and Figure \ref{fig:breakdown_1500} represent the detailed elapsed time for processing 100 and 1500 AC requests with \textit{eXACML+}, respectively. We can see that time taken to make access control decisions and to manipulate query graphs take less than 0.01 second in all requests and are rather consistent despite the increment number of requests and loaded policies, which are 50 for \ref{fig:breakdown_100} and 1000 for \ref{fig:breakdown_1500}. Response time of sending query graphs to DSMS occupies one third of the total response time on average and has much larger variance. We notice that there are a few cases where sending query graphs to DSMS take much longer time than average and these cases take place only in the beginning of the request sequences. We believe this is due to the behaviour of StreamBase API, which needs longer time to establish initial connections to StreamBase than to send subsequence queries. In general, the response time for \textit{eXACML+} to process AC requests is consistent for over 99\% of the requests, which verifies that the whole system is scalable with respect to number of requests processed and number of polices loaded

Figure \ref{fig:zipf_dis} shows how the overhead changes when the proxy enables caching. Although \textit{eXACML+} does not outperform direct query systems, the performance improvement brought by caching is substantial. Unlike \textit{eXACML}, what cached in the proxy is not actual data, but data stream handles, whose sizes are significantly smaller.  The performance improvement thus is not as obvious as in \cite{anh12}. Nevertheless, caching leads to over 100\% improvement over non-cached requests for nearly 40\%  of the number of quests and at least 10\% improvement for the rest requests. The results justify the importance to have cache mechanism implemented in proxy when the request distribution is heavy-tailed.

\section{Related Work}
\label{sec:relatedwork}

There are a couple of cloud-based systems that aim to provide data sharing capabilities across the Internet. Dropbox\cite{dropbox} and iCloud\cite{icloud} are examples of commercial products that enables file sharing among individual devices bases on their cloud storage back-end. SenseWeb\cite{senseweb} and SensorBase\cite{sensorbase}, on the other hand, allows users to upload the share their sensor data. Theses systems support coarse-grained access control model in which an user either makes his data public, shares with a list of people or just keeps it private. They cannot deal with the access control scenarios we considered in this paper and in \cite{anh12}.

In recently years, Time-series data, or stream data management systems have been developed substantially. So far the most renowned commercial DSMS, StreamBase \cite{streambase}, was evolved from famous Aurora system\cite{abadi03} and its distributed version borealis\cite{borealis}. Enforce access control on data streams is still a fresh topic in the research community. Carminati et al. \cite{carminati07}, \cite{carminati07a} proposed model and framework for enforcing access control over stream data. Rimma, et al. \cite{rimma08}, \cite{rimma09} proposed to make use of embedded punctuations in the data stream to enforce access control, while Lindner. et al.\cite{lindner05}, \cite{lindner06} proposed to build an additional static layer on top of query engines. All these approaches are built on  coarse-grained access control models and lack of capability to deal with the scenario where fine-grained policies are needed.

Using of XACML\cite{oasisxacml} in Cloud systems is yet to receive further development. \cite{power05} uses XACML in grid environment where its role is to unify database access control mechanisms from multiple parties.
We foresee that XACML will be more widely used in the industry as it's the current OASIS standard and is continuously evolving in response to new access control requirements. To the best of our knowledge, our work in this paper is the first to use XACML model to enforce fine-grained access control policies over stream data.

\section{Conclusion and Future Work}
\label{sec:conclusion}

In this paper, we have proposed \textit{eXACML+} facilitating data stream owners to share their data streams with other users in a secure and flexible manner over a trusted cloud infrastructure. The main challenges were owing to the vital differences between bounded data (as in RDBMS) and unbounded stream data (DSMS). We also explore the problem of possible privacy leak by allowing a single user to have access to multiple aggregated streams of one master data stream. We also show in the paper how to effectively detect if a user query will return empty or only partial result due to mismatches with access control policies. We have implemented the prototype eXACML+ system over the Aurora data-model based StreamBase data stream management system. Preliminary experiment results show our framework's efficacy. It incurs constant and consistent overhead compared with direct queries (without access control) on the data stream engine.


Our immediate plans are to migrate the framework to commercial Cloud environments such as Amazon EC2 \cite{ec2} and Microsoft's Azure \cite{azure} for more comprehensive evaluations with practical workloads instead of synthetic ones. Moreover, we intend to use other stream base engine like APE \cite{ape} and DB2 DSE \cite{db2dse} to broaden the range of applications that our framework supports. On the conceptual front, relaxing the trusted cloud model to incorporate more accountability mechanisms is our primary next challenge. 

\bibliographystyle{splncs}
\bibliography{References}

\end{document}